\documentclass[12pt,epsfig]{article}
\usepackage[xdvi]{graphicx}

\newcommand{\beq}{\begin{equation}}
\newcommand{\eeq}{\end{equation}}
\newcommand{\bea}{\begin{eqnarray}}
\newcommand{\eea}{\end{eqnarray}}

\def\e2sig{e^{-2r\sigma}}

\begin{document}
\setlength{\baselineskip}{0.7cm}
\begin{titlepage} 
%%%%% PREPRINT NUMBERS %%%%%%
\begin{flushright}
%KOBE-TH-07-11 \\
OCU-PHYS-378
%KEK-TH-1205 
\end{flushright}
\vspace*{10mm}
%%%%%%%%%%%%%%%%%%% TITLE %%%%%%%%%%%%%%%%%%
\begin{center}{\Large \bf 
Diphoton and $Z$ photon Decays of Higgs Boson 
\vspace*{3mm}
in Gauge-Higgs Unification: 
A Snowmass white paper
}
\end{center}
%%%%%%%%%%%%%%%% AUTHORS %%%%%%%%%%%%%%%%%%%%%%%
\vspace*{10mm}
\begin{center}
{\Large Nobuhito Maru}$^{a}$ and {\Large Nobuchika Okada}$^{b}$
\end{center}
%%%%%%%%%%%%%%%%%%%%%%% AFFILIATION %%%%%%%%%%%%
\vspace*{0.2cm}
\begin{center}
%\small
${}^{a}${\it Department of Mathematics and Physics, Osaka City University, \\
Osaka 558-8585, Japan}
\\[0.2cm]
${}^{b}${\it 
Department of Physics and Astronomy, University of Alabama, \\
Tuscaloosa, Alabama 35487, USA} 
%%%%%
\end{center}
%%%%%%%%%%%%%%%%%% ABSTRACT %%%%%%%%%%%%%%%
\vspace*{2cm}
\begin{abstract}

In the context of gauge-Higgs unification scenario  
 in a 5-dimensional flat spacetime, 
 we investigate Higgs boson production via gluon fusion 
 and its diphoton and $Z\gamma$ decay modes at the LHC. 
We show that the signal strength of the Higgs diphoton decay mode 
 observed at ATLAS (and CMS), which is considerably larger 
 than the Standard Model expectation, 
 can be explained by a simple gauge-Higgs unification model 
 with color-singlet bulk fermions to which  
 a half-periodic boundary condition is assigned. 
The bulk fermions with mass at the TeV scale 
 also play a crucial role in reproducing 
 the observed Higgs boson mass of around 125 GeV. 
One naturally expects that the KK modes also 
 contribute to the effective $H-Z-\gamma$ coupling. 
However, we show a very specific and general prediction 
 of the gauge-Higgs unification scenario 
 that KK-mode contributions to the $H-Z-\gamma$ coupling 
 do not exist at the 1-loop level. 
If the excess of the Higgs to diphoton decay mode persists, 
 its correlation with the Higgs to $Z$ photon decay mode 
 can be a clue to distinguish scenarios beyond the SM, 
 providing a significant improvement of the sensitivity 
 for the Higgs boson signals in the future.

\end{abstract}
\end{titlepage}
%%%%%%%%%%

%%%%%%%%%%%%%%%%%%%%%%
%\section{Introduction}
%%%%%%%%%%%%%%%%%%%%%%
As announced on July 4th 2012, the long-sought Higgs boson 
 was finally discovered by ATLAS \cite{ATLAS} and CMS \cite{CMS} 
 collaborations at the Large Hadron Collider. 
The discovery is based on the Higgs boson search 
 with a variety of Higgs boson decay modes. 
Although the observed data were mostly consistent with 
 the Standard Model (SM) expectations, 
 the diphoton decay mode showed the signal strength 
 considerably larger than the SM prediction. 
Since the effective Higgs-to-diphoton coupling is induced 
 at the quantum level even in the SM, a certain new physics 
 can significantly affect the coupling. 
%This fact motivated many recent studies 
% for a possible explanation of the excess 
% in the Higgs to diphoton decay mode by various extensions 
% of the SM with supersymmetry~\cite{AHC-SUSY} or 
% without supersymmetry~\cite{AHC-NonSUSY, MO2}. 
Although the updated CMS analysis~\cite{CMS2} gives 
 a much lower value for the signal strength 
 of the diphoton events than the previous one, 
 the updated ATLAS analysis~\cite{ATLAS2} is still consistent 
 with their earlier result. 
The excess may persist in future updates.

Gauge-Higgs unification (GHU) \cite{GH} is 
 one of the fascinating scenarios for physics beyond the SM, 
 which can provide us a solution to the gauge 
 hierarchy problem without invoking supersymmetry. 
In this scenario, the SM Higgs doublet is identified 
 with an extra spatial component of a gauge field 
 in higher dimensional gauge theory. 
Nevertheless the scenario is non-renormalizable, 
 the higher dimensional gauge symmetry 
 allows us to predict various finite physical observables 
 such as Higgs potential~\cite{Higgsmass}, 
 $H\to gg, \gamma\gamma$~\cite{MO, Maru}, 
 the electric and magnetic moment of fermion~\cite{g-2EDM}. 

%%%%%%%%%%%%%%%%%%%%%%%%%%%%%%%%%%%%%%%%%%%%%%%%
%\section{Higgs production and diphoton decay in GHU}
%%%%%%%%%%%%%%%%%%%%%%%%%%%%%%%%%%%%%%%%%%%%%%%%
We consider a simple GHU model based on the gauge group 
 $SU(3) \times U(1)'$ in a 5-dimensional flat space-time 
 with orbifolding on $S^1/Z_2$ with radius $R$ of $S^1$. 
In our setup of bulk fermions, we follow Ref.~\cite{CCP}: 
 the up-type quarks except for the top quark, 
 the down-type quarks and the leptons are embedded 
 respectively into ${\bf 3}$, $\overline{{\bf 6}}$, 
 and ${\bf 10}$ representations of $SU(3)$. 
In order to realize the large top Yukawa coupling, 
 the top quark is embedded into a rank $4$ representation 
 of $SU(3)$, namely $\overline{{\bf 15}}$. 
The extra $U(1)'$ symmetry works to yield 
 the correct Weinberg angle~\cite{SSS}. 
 %and the SM $U(1)_Y$ gauge boson 
 %is given by a linear combination between the gauge bosons 
 %of the $U(1)'$ and the $U(1)$ subgroup in $SU(3)$ \cite{SSS}
% \footnote{
% It is known that the correct Weinberg angle can also be obtained 
% by introducing brane localized gauge kinetic terms \cite{SSS}, 
% but we do not take this approach in this paper.}.  
We assign appropriate $U(1)'$ charges for bulk fermions 
 to give the correct hyper-charges for the SM fermions.

The boundary conditions should be suitably assigned 
 to reproduce the SM fields as the zero modes. 
While a periodic boundary condition corresponding to $S^1$ 
 is taken for all of the bulk SM fields, 
 the $Z_2$ parity is assigned for gauge fields and fermions 
 in the representation ${\cal R}$ 
 by using the parity matrix $P={\rm diag}(-,-,+)$ as
\bea
A_\mu (-y) = P^\dag A_\mu(y) P, \quad A_y(-y) =- P^\dag A_\mu(y) P,  \quad 
\psi(-y) = {\cal R}(P)\psi(y) 
\label{parity}
\eea 
 where the subscripts $\mu$ ($y$) denotes the four (the fifth) 
 dimensional component. % and $R$ is a radius of $S^1$. 
With this choice of parities, the $SU(3)$ gauge symmetry is 
 explicitly broken to $SU(2) \times U(1)$. 
A hypercharge is a linear combination of $U(1)$ and $U(1)'$.  
% in this setup. 
%One may think that the $U(1)_X$ gauge boson which is 
% orthogonal to the hypercharge $U(1)_Y$ also has a zero mode. 
%However, the $U(1)_X$ symmetry is anomalous in general 
% and broken at the cutoff scale and hence,  
% the $U(1)_X$ gauge boson has a mass of order 
% of the cutoff scale \cite{SSS}. 
As a result, zero-mode vector bosons in the model are 
 only the SM gauge fields.

Off-diagonal blocks in $A_y$ have zero modes 
 because of the overall sign in Eq.~(\ref{parity}), 
 which corresponds to an $SU(2)$ doublet. 
In fact, the SM Higgs doublet ($H$) is identified as 
\bea
A_y^{(0)} = \frac{1}{\sqrt{2}}
\left(
\begin{array}{cc}
0 & H \\
H^\dag & 0 \\
\end{array}
\right). 
\eea
The KK modes of $A_y$ are eaten by KK modes of the SM gauge bosons 
 and enjoy their longitudinal degrees of freedom 
 like the usual Higgs mechanism.

The parity assignment also provides the SM fermions as massless modes, 
 but it also leaves exotic fermions massless. 
Such exotic fermions are made massive by introducing 
 brane localized fermions with conjugate $SU(2) \times U(1)$ charges 
 and an opposite chirality to the exotic fermions, 
 allowing us to write mass terms on the orbifold fixed points. 
In the GHU scenario, the Yukawa interaction is unified 
 with the gauge interaction, so that the SM fermions 
 obtain the mass of the order of the $W$-boson mass 
 after the electroweak symmetry breaking. 
To realize light SM fermion masses, one may introduce 
 a $Z_2$-parity odd bulk mass terms for the SM fermions, 
 except for the top quark. 
Then, zero mode fermion wave functions with opposite chirality 
 are localized towards the opposite orbifold fixed points 
 and a resulting Yukawa coupling is exponentially 
 suppressed by the overlap integral of the wave functions. 
%In this way, all exotic fermion zero modes become heavy 
% and small Yukawa couplings for light SM fermions 
% are realized by adjusting the bulk mass parameters. 
In order to realize the top quark Yukawa coupling, 
 we introduce a rank $4$ tensor representation, namely, 
 a symmetric $\overline{{\bf 15}}$ without a bulk mass \cite{CCP}, 
 which leads to  
%This leads to a group theoretical factor $2$ enhancement 
% of the top quark mass as 
 $m_t = 2 m_W$ at the compactification scale \cite{SSS}. 
%Note that this mass relation is desirable 
% since the top quark pole mass receives QCD threshold 
% corrections which push up the mass about 10 GeV. 
%See, for example, Ref.~\cite{FlavorCP} 
% for flavor mixing and CP violation in the GHU scenario. 

\begin{figure}[htbp]
  \begin{center}
   \includegraphics[width=80mm]{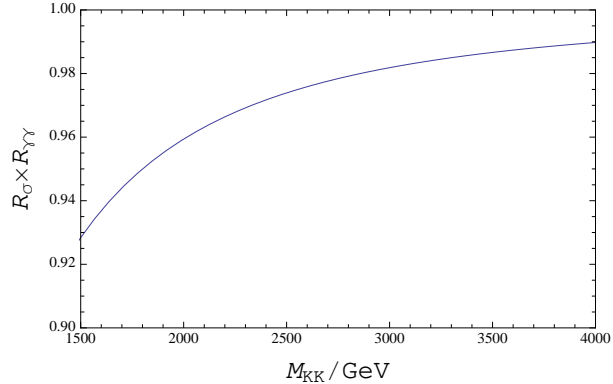}
      \end{center}
  \caption{
 The ratio of diphoton events in the simple GHU model  
 to those in the SM as a function of the compactification scale.}
\label{diphotonSM}
\end{figure}

%%%%%%%%%%%%%%%%%%%%%%%%%%%%%%%%%%%%%%%%%%%%%%%
%\subsection{$gg \to h \to \gamma\gamma$}
%%%%%%%%%%%%%%%%%%%%%%%%%%%%%%%%%%%%%%%%%%%%%%%
With the setup discussed above, 
 we have estimated the ratio of 
 the signal strength of the process 
 $gg \to H \to \gamma \gamma$ in our model 
 to the one in the SM \cite{MO, MO2}. 
%Putting all together, we find 
%\bea
%{\rm R} \equiv \frac{\sigma(gg \to h \to \gamma \gamma)
%}{\sigma(gg \to h \to \gamma \gamma)_{\rm SM}}
%=R_\sigma \times R_{\gamma\gamma} 
%\simeq 1 - \frac{374}{141} \pi^2  \left(\frac{m_W}{m_{\rm KK}} \right)^2.  
%\eea
The result is depicted in Fig.~\ref{diphotonSM} 
 as a function of the KK mode mass/the compactification scale. 
The ratio R is found to be smaller than one, 
 because of the destructive KK mode contribution 
 to the gluon fusion channel and the accidental 
 cancellation among the KK mode contributions to 
 the Higgs-to-diphoton decay width. 
This fact has already been advocated in the previous paper 
 by the present authors \cite{MO}.

%%%%%%%%%%%%%%%%%%%%%%%%%%%%%%%%%%
%  Fig
%%%%%%%%%%%%%%%%%%%%%%%%%%%%%%%%%%
%%%%%%%%%%%%%%%%%%%%%%%%%%%%%%%%%%

\begin{figure}[htbp]
  \begin{center}
   \includegraphics[width=70mm]{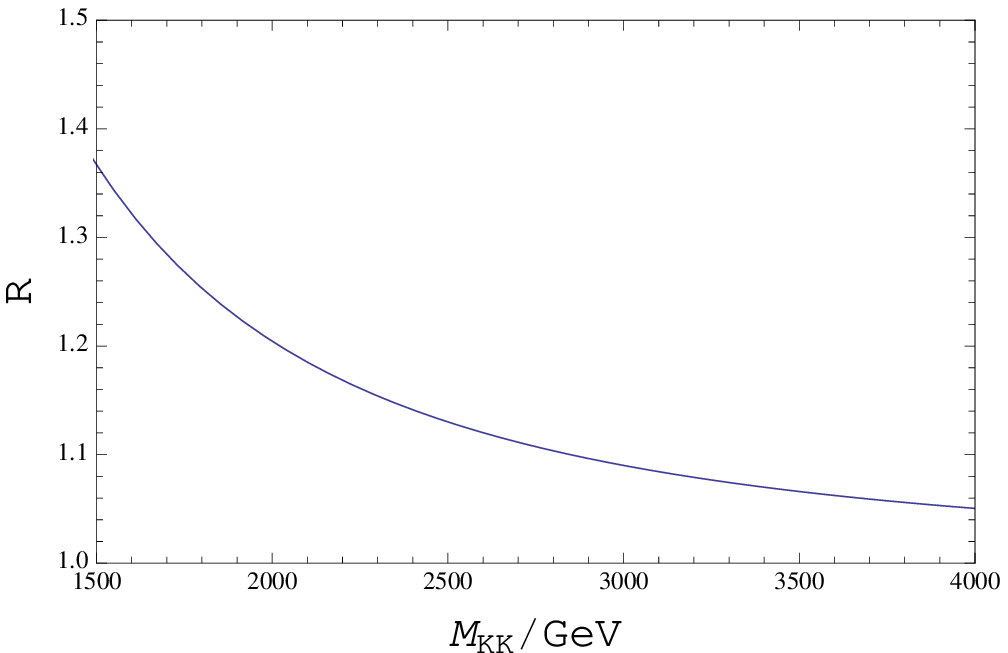}
   \hspace*{10mm}
   \includegraphics[width=70mm]{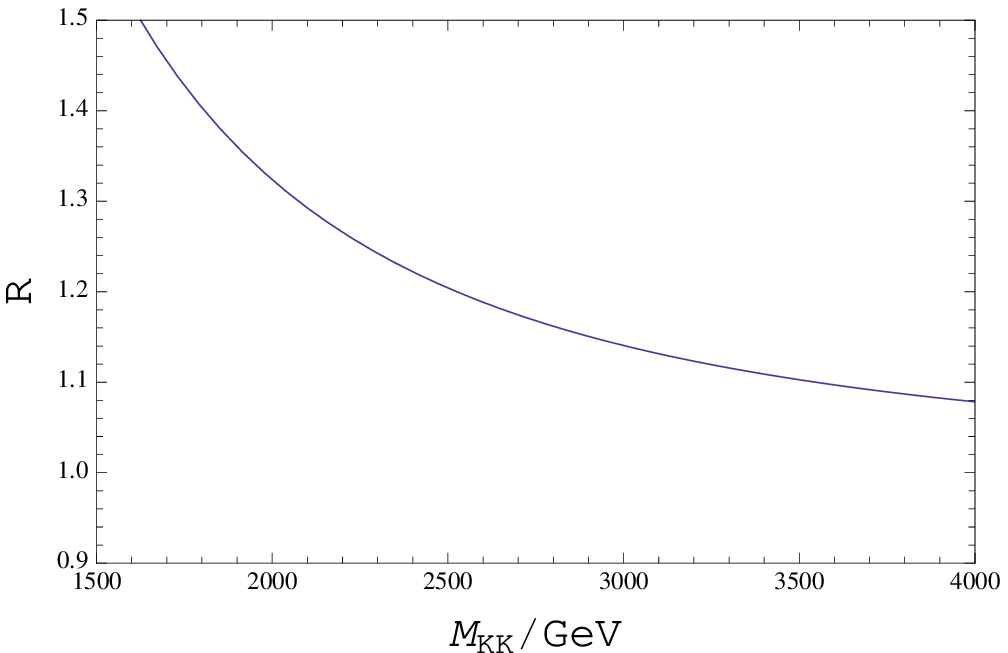}
   \end{center}
  \caption{
The diphoton signal strength (normalized by the SM prediction) 
 in the GHU model with the ${\bf 10}$-plet (left) 
 and ${\bf 15}$-plet (right) bulk fermions
 as a function of the compactification scale.
%Here we have used $Q=-1$ and $c_B=0.23$ for 
% the left panel, while $Q=-5$ and $c_B=0.69$ for the right panel.
}
  \label{diphotonwithleptonR}
\end{figure}

Now we extend the present GHU model to account 
 for the signal strength measured by ATLAS (and CMS) 
 for the process $gg \to H \to \gamma \gamma$ 
 which is considerably larger than the SM expectation. 
The simplest extension is to introduce color-singlet bulk 
 fermions with the half-periodic boundary condition in the bulk.  
% $\psi(y+2\pi R) = - \psi(y)$, in the bulk. 
%The main reasons for this strategy are two folds. 
%The first is that since the KK mode fermion contribution 
% is destructive to the SM fermion contribution, 
% colored KK mode contribution is not desirable 
% for the Higgs production process via gluon fusion. 
%On the other hand, the KK mode contribution is 
% constructive to the SM one for the Higgs-to-diphoton couple. 
%Thus, the introduction of color-singlet bulk fermions 
% nicely work to enhance the diphoton signal strength. 
%The second is that the half-periodic bulk fermion has 
% no massless mode, and unwanted exotic massless fermions 
% do not come out in the model. 
%Another advantage of the half-periodic bulk fermion is 
% that its first KK mode mass is smaller than the compactification 
% scale and its loop corrections dominate over 
% those from the KK modes of the periodic bulk fermion. 
%Furthermore, the existence of the half-periodic bulk fermion 
% is crucial to achieve a Higgs boson mass around 125 GeV 
% in our GHU model, as we will discuss in the next section. 
In \cite{MO2}, we have considered two examples for 
 the color-singlet bulk fermions of the representations  
 ${\bf 10}$ and ${\bf 15}$ of $SU(3)$, 
 with a suitable $U(1)'$ charge $Q$ assignment 
 and bulk mass $M$ parametrized in the unit of the KK mass $m_{\rm KK}=1/R$ 
 by $c_B \equiv M/m_{\rm KK}$. 
For the two cases, we plot the ratio R as a function of  
 the KK mode mass $m_{\rm KK}$ in Fig.~\ref{diphotonwithleptonR}. 
The left(right) panel corresponds to the case 
 with the ${\bf 10}({\bf 15})$-plet bulk fermion, 
 where we have fixed $Q=-1(-5)$ and $c_B=0.23(0.69)$. 
As we will see later, 
 the Higgs boson mass around 125 GeV
 can be reproduced with the bulk mass $c_B=0.23(0.69)$ 
 for $m_{\rm KK}=3$ TeV. 
%The result for the case with the ${\bf 15}$-plet bulk fermion 
 %is depicted in the right panel for $Q=-5$ and $c_B=0.69$. 
%This bulk mass reproduces the Higgs boson mass around 125 GeV. 
We have found that the Higgs-to-diphoton signal strength is 
 considerably enhanced in the presence of 
 the half-periodic bulk fermions with the TeV scale mass. 
The rate of the enhancement can be large as we like by adjusting a $U(1)'$ charge $Q$.

%%%%%%%%%%%%%%%%%%%%%%%%%%%%%%%%%%%%%%%%%%%%%%%%%%%%%
% Fig 
%%%%%%%%%%%%%%%%%%%%%%%%%%%%%%%%%%%%%%%%%%%%%%%%%%%%%
%%%%%%%%%%%%%%%%%%%%%%%%%%%%%%%%%%%%%%%%%%%%%%%%%%%%%

\begin{figure}[htbp]
  \begin{center}
   \includegraphics[width=70mm]{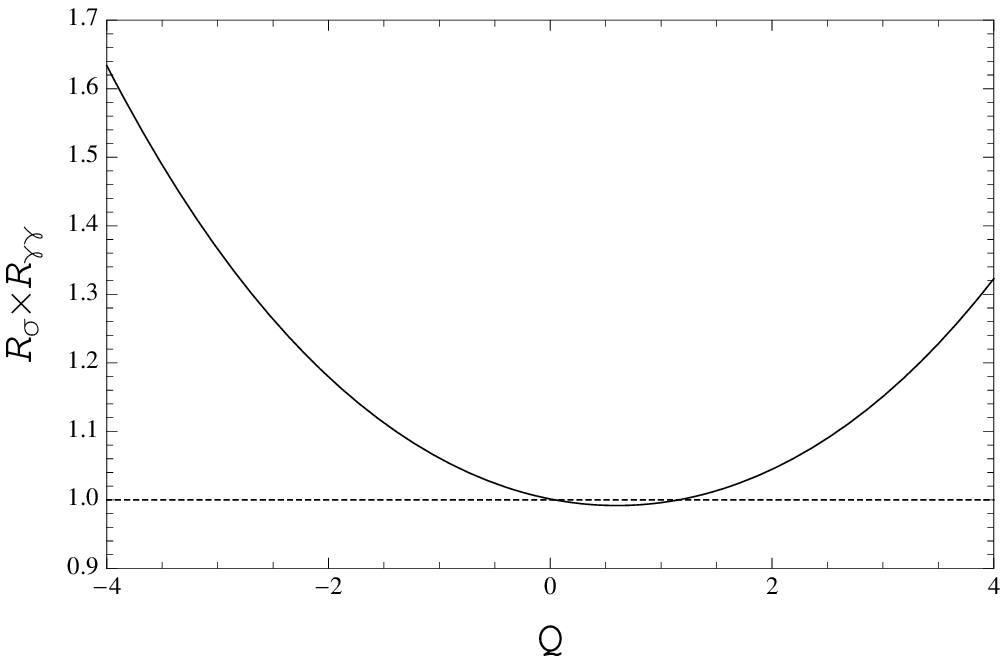}
   \hspace*{10mm}
   \includegraphics[width=70mm]{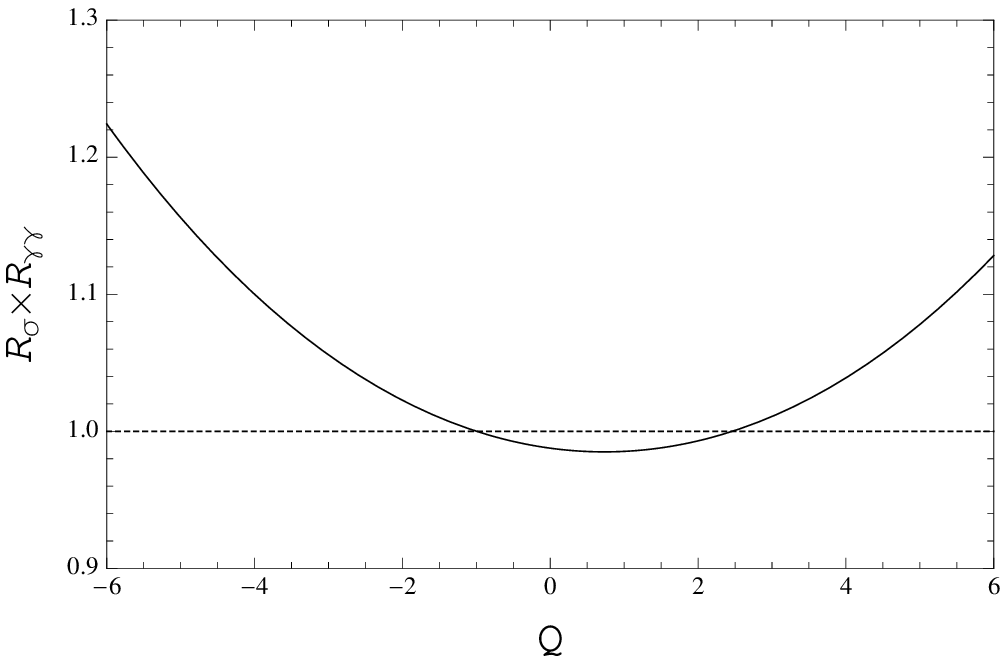}
   \end{center}
  \caption{
The diphoton signal strength (normalized by the SM prediction) 
 in the GHU model with the ${\bf 10}$-plet (left) 
 and ${\bf 15}$-plet (right) bulk fermions
 as a function of the $U(1)'$ charge $Q$, 
 for $m_{\rm KK}=3$ TeV. 
%The bulk masses are fixed as $c_B=0.23$ for 
% the left panel, while $c_B=0.69$ for the right panel.
}
 \label{diphotonwithleptonQ}
\end{figure}

In Fig.~\ref{diphotonwithleptonQ}, 
 we plot the ratio of diphoton signal strength to 
 the SM one as a function of the $U(1)'$ charge $Q$, 
 for the two cases. 
For each plot, the bulk masses are fixed 
 to be the same values as in the previous plots. 
We can see that $|Q|={\cal O}(1)$ is enough 
 to give rise to an order 10\% enhancement 
 of the diphoton signal. 
%In general, a larger representation field 
% leads to more enhancements than those by smaller 
% representations, since large representations 
% include more fields with higher $U(1)$ charges 
% in the SM decomposition. 
%In Fig.~\ref{diphotonwithleptonQ}, 
% the deviation of the diphoton signal strength 
% for the case with the ${\bf 15}$-plet fermion  
% is milder than the case with the ${\bf 10}$-plet. 
%This is because  a large bulk mass is assigned 
% for the case with the ${\bf 15}$-plet fermion
% and the KK modes are heavier. 

%%%%%%%%%%%%%%%%%%%%%%%%%%%%%%%%%%%%%%%%%%%%%%%%%%%%%
% Fig 
%%%%%%%%%%%%%%%%%%%%%%%%%%%%%%%%%%%%%%%%%%%%%%%%%%%%%
%%%%%%%%%%%%%%%%%%%%%%%%%%%%%%%%%%%%%%%%%%%%%%%%%%%%%

%%%%%%%%%%%%%%%%%%%%%%%%%%%%%%%%%%%%%%%%%%%%
%\section{Estimate of Higgs boson mass}
%%%%%%%%%%%%%%%%%%%%%%%%%%%%%%%%%%%%%%%%%%%%
Next, we discuss how the Higgs boson mass 
 around 125 GeV is realized in our model. 
Realizing the 125 GeV Higgs boson mass 
 %as well as 
 %the electroweak symmetry breaking 
 is a quite non-trivial %phenomenological issue 
 in 5-dimensional GHU scenario 
%This is because the Higgs doublet is embedded 
% in the five dimensional component  of the bulk gauge  
% field and as a result, the Higgs doublet has 
% no scalar potential at the tree level . 
%The electroweak symmetry should be broken dynamically, 
% in other words, at the quantum  level. 
 since the Higgs quartic coupling is 
 generated at loop levels and a calculated Higgs boson mass is likely to be small. 
%Towards a realistic GHU scenario, a variety of extra bulk fields 
% with suitable boundary conditions and bulk/brane mass terms 
% have been considered (see, for example, Refs.~\cite{SSS, CCP}). 
%It is therefore a highly non-trivial task to propose 
% a simple and phenomenologically viable GHU model. 
In estimating Higgs boson mass, 
 we take a 4-dimensional effective theory approach 
 developed by Ref.~\cite{GHcondition}, 
 in which
%As has been shown in this paper, 
 the low energy effective theory of the 5-dimensional GHU scenario 
 is equivalent to the SM with the so-called ``gauge-Higgs condition'' 
 on the Higgs quartic coupling, 
 namely, we impose a vanishing Higgs quartic coupling
 at the compactification scale, which reflects the 5-dimensional gauge 
 invariance restoration.  
 %at an energy higher than the compactification scale. 
%Employing this effective theory approach, 
 The Higgs boson mass at low energies is easily calculated by solving the RGE 
 of the Higgs quartic coupling with the gauge-Higgs condition and  
% instead of calculating the Coleman-Weinberg potential 
% of the Higgs doublet. 
%We assume that the electroweak symmetry breaking correctly occurs  
% by the introduction of a suitable set of bulk fermions. 
%Note that the effective Higgs mass squared is quadratically 
% sensitive to the mass of heavy states, 
% while the effective Higgs quartic coupling is dominantly  
% determined by interactions of the Higgs doublet with light states. 
%Therefore, the Higgs boson mass at low energies 
 is mainly determined by light states below the compactification scale. 
% once we assume the correct electroweak symmetry breaking. 
In our model, we have introduced bulk fermions 
 with the half-periodic boundary condition, 
 and their first KK modes appear 
 below the compactification scale. 
Therefore, not only the SM particles but also 
 the first KK modes are involved in our RGE analysis 
 with the gauge-Higgs condition. 

%%%%%%%%%%%%%%%%%%%%%%%%%%%%%%%%%%%%%%%%%%%%%%%%%%%%%
% Fig 
%%%%%%%%%%%%%%%%%%%%%%%%%%%%%%%%%%%%%%%%%%%%%%%%%%%%%
\begin{figure}[htbp]
  \begin{center}
   \includegraphics[width=90mm]{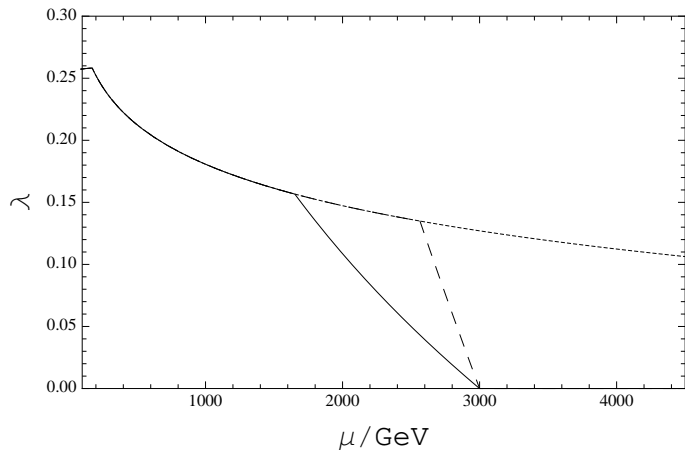}
  \end{center}
  \caption{
1-loop RGE running of the Higgs quartic coupling. 
%The solid (dashed) line corresponds  to the case 
% of the ${\bf 10}$-plet (${\bf 15}$-plet) bulk fermion 
% with the bulk mass $c_B=0.23$ ($c_B=0.69$). 
%Here the compactification scale is fixed 
% as $m_{\rm KK}=1/R = 3$ TeV, at which 
% the gauge-Higgs condition ($\lambda(m_{\rm KK})=0$) 
% is applied. 
%The dotted line shows the running of 
% the SM Higgs quartic coupling 
% with the boundary condition,
% $\lambda(\mu=m_h)=0.258$, corresponding 
% to the Higgs pole mass $m_h=125$ GeV. 
}
  \label{RGElambda}
\end{figure}
%%%%%%%%%%%%%%%%%%%%%%%%%%%%%%%%%%%%%%%%%%%%%%%%%%%%%

The numerical results of 1-loop RGE of the Higgs quartic coupling 
 are shown in Fig.~\ref{RGElambda}. 
Here we have applied the gauge-Higgs condition ($\lambda(m_{\rm KK})=0$)
 at the compactification scale $m_{\rm KK}=3$ TeV 
 and numerically solve the RGE toward low energies. 
%In the analysis, we have used 
% $y_t(\mu) =0.943$ for $\mu \geq m_t=173.1$ GeV 
% ($y_t(\mu)=0$ for $\mu < m_t=173.1$ GeV), 
% and $g_1=0.459$, and $g_2=0.649$ at the $Z$-boson mass scale. 
%For simplicity, we estimate the Higgs boson pole mass 
% by the condition $ \lambda(\mu=m_h) v^2 =m_h^2$. 
%In Fig.~\ref{RGElambda}, 
The bulk masses of the ${\bf 10}({\bf 15})$-plet 
 are fixed to be the values $c_B=0.23(0.69)$, respectively, 
 with which Higgs boson mass of $m_H=125$ GeV 
 (equivalently, $\lambda (\mu=m_H) = 0.258$) is realized. 
The solid (dashed) line represents the running Higgs quartic 
 coupling for the case with the ${\bf 10}({\bf 15})$-plet bulk fermion, 
 while the dotted line corresponds to the RGE running 
 in the SM case with the boundary condition $\lambda (\mu=m_H) = 0.258$. 
%In this rough analysis, the Higgs quartic coupling becomes zero 
% at $\mu \sim 10^{4.5}$ GeV. 
As can be seen from Fig.~\ref{RGElambda},
 the existence of the half-periodic bulk fermions 
 is essential to realize the Higgs mass around 125 GeV 
 with the compactification at the TeV scale. 
Since the bulk fermions provide many first KK mode 
 fermions in the SM decomposition, 
 the running Higgs quartic coupling is sharply 
 rising from zero toward low energies. 
%In addition, the bulk mass also plays a crucial role 
% to adjust the resultant Higgs boson mass to be 125 GeV. 

We naturally expect that the decay $H \to Z \gamma$ is also deviated from the SM prediction 
 since the KK modes have electroweak charges. 
The correlation between the $\gamma\gamma$ and 
 the $Z \gamma$ decays of Higgs boson is interesting 
 since this property is model dependent and 
 useful for distinguishing our model from other models beyond the SM. 
In \cite{MO3}, we have studied the KK mode contributions to 
 the Higgs boson to $Z \gamma$ decay in the GHU  
 and found a striking result that we have no KK mode contributions 
 to it the at 1-loop level. 
This is because $Z$ boson always has couplings 
 over two different mass eigenstates corresponding 
 to the mass splitting due to the electroweak symmetry breaking, 
 while the Higgs boson and photon couple with the same mass
 eigenstates. 
This coupling manner originates from the basic structure 
 of the GHU scenario and can be a clue to distinguish 
 the GHU scenario from other scenarios beyond the SM, 
 providing a significant improvement of the sensitivity 
 for the Higgs boson signals in the future.

%%%%%%%%%%%%%%%%%%%%%%%%%%%%
\subsection*{Acknowledgment}
%%%%%%%%%%%%%%%%%%%%%%%%%%%%
The work of N.M. is supported in part by the Grant-in-Aid 
 for Scientific Research from the Ministry of Education, 
 Science and Culture, Japan No. 24540283.
The work of N.O. is supported in part 
 by the DOE Grant No. DE-FG02-10ER41714.

\
\end{document}